\documentclass[11pt]{article}
\usepackage{moriond}

\usepackage{amsmath} 
\usepackage{txfonts} 

\def\Journal#1#2#3#4{{#1} {\bf #2}, #3 (#4)}

\def\be{\begin{equation}}
\def\ee{\end{equation}}
\def\bea{\begin{eqnarray}}
\def\eea{\end{eqnarray}}

\newcommand{\Photo}{}

\begin{document}
\vspace*{4cm}
\title{Underground Commissioning of LUX}

\author{Michael Woods, for the LUX Collaboration}

\address{Department of Physics, One Shields Avenue,\\
Davis, CA 95616, USA}

\maketitle\abstracts{
LUX is a dual-phase xenon TPC designed for the direct detection of dark matter. Using 370~kg of xenon, LUX is capable of setting a WIMP-nucleon cross section limit at $2 \times 10^{-46}\mbox{~cm}^2$ after 300 days of running. LUX will surpass all existing dark matter limits for WIMP masses above 10~GeV within weeks of beginning its science run. Following a successful six month surface run, the detector has recently been deployed underground, and we expect completed commission in the near future. Updates on status and results are provided.}

\section{Introduction}

Evidence in favor of a mass component of the universe that is uncoupled from the generation of electromagnetic radiation has been accumulating for 80 years~\cite{zw}. It is  known that non-baryonic dark matter makes up a majority of the mass density of the universe~\cite{ja}. Observational evidence is plentiful and includes galaxy cluster rotation profiles that suggest large, extended mass distributions that conflict with the mass distribution of the largest known mass contributor, intergalactic gas. A weakly interacting massive particle (WIMP) can be constructed in a number of models{~\cite{ju} as a heavy dark matter candidate that would interact with baryonic matter via nuclear scattering~\cite{go}.

\section{Operating Principle}

The Large Underground Xenon Experiment (LUX) has deployed a dual phase xenon time projection chamber for direct detection of dark matter. Liquid xenon's density makes it a very attractive target for direct dark matter searches.  The LUX detector contains 370~kg of liquid xenon with 300~kg in the active region where signal can be extracted. To illustrate our event topology, an event schematic in LUX is shown in Figure~\ref{fig:event}. Energy depositions are partitioned in to scintillation light and ionization. Immediate partial recombination of the ionization signal with Xe ions can occur leading to additional scintillation. The scintillation signal (S1 signal) is prompt and collected in timescales on the order of 100 ns.

An electric field is established from the top of the gaseous xenon to the bottom of the liquid xenon space. Ionized electrons are slowly drifted to the liquid surface where electron extraction occurs. Electrons enter into the gaseous region where acceleration in the electric field causes electroluminescence, creating additional scintillation light in the timespan of microseconds. This delayed ionization signal (S2 signal) is separated in time from the S1 signal proportional to the depth of the energy deposition, and measurement of the time difference allows realization of the $z$ location of the event. The S2 signal provides $(x,y)$ information thus permitting full three-dimensional position reconstruction.

\begin{figure}
\begin{center}
\begin{minipage}{0.33\linewidth}
\centerline{\includegraphics[width=1.3\linewidth,draft=false]{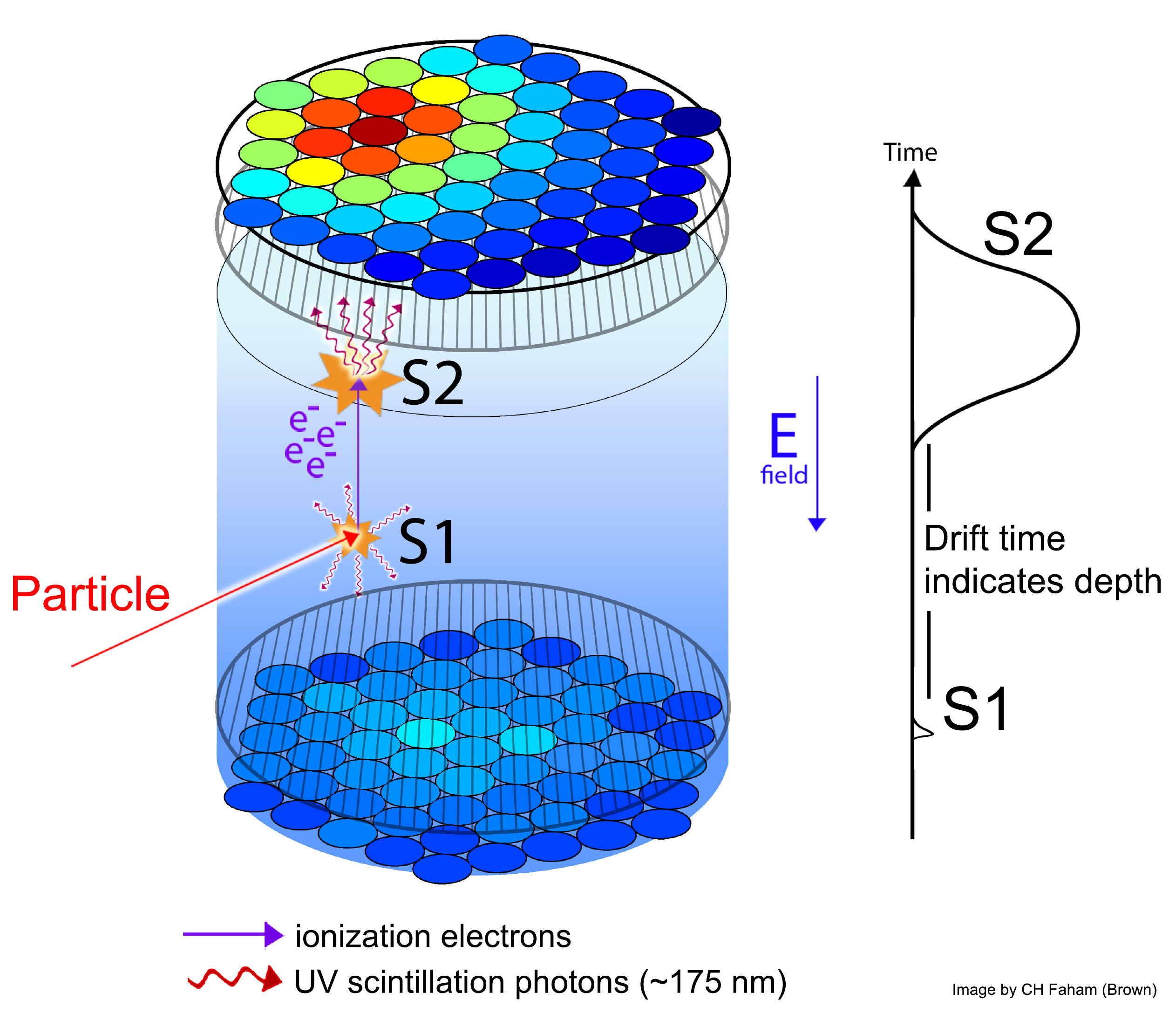}}
\end{minipage}
\caption{Event schematic in the LUX detector. An energy deposition causes ionization and prompt light (S1). An applied electric field drifts ionization electrons to the gas gap where extraction occurs and electroluminescence creates a delayed light signal (S2) . The size of the S2 is proportional to the charge that reached the surface, and the time separation is proportional to the depth of the energy deposition in the liquid xenon.}
\label{fig:event}
\end{center}
\end{figure}

\section{The Detector}
The details of the detector are described in greater detail elsewhere~\cite{ak_lux}. An inner and outer cryostat constructed from low-background titanium~\cite{ak_ti} house the internals of the experiment. Internal support structures that are not load-bearing  are machined oxygen-free high thermal conductivity copper.

 LUX is instrumented with 122 low-background Hamamatsu R8778 PMTs~\cite{ak_pmt}. The 2'' tubes have a measured quantum efficiency $\sim$33\% and are separated in to two banks of 61 tubes. The upper bank reconstructs $(x, y)$ position based on the hit pattern of the S2 signal in gaseous xenon while the bottom array is located in liquid xenon and benefits from enhanced light collection for the primary scintillation signal providing improved energy reconstruction. 

The electric field is separated into multiple field regions by a collection of four wire grids and one wire mesh.  Near the top and bottom PMT banks, reverse biased grids create electromagnetic shields to prevent charge collection from occurring around the photosensors. The cathode grid at the bottom of the active region delivers a large negative voltage that establishes a drift field across the bulk of the detector. A gate grid just below the liquid surface marks the end of the drift field and the beginning of the extraction field. A large positive voltage applied to the anode mesh in the gas phase, together with the gate grid, create (over a distance of 1~cm) a large extraction field for removing electrons from the liquid.
 
\section{Surface Run}

The LUX detector was deployed in a surface laboratory at the Sanford Underground Research Facility (SURF) in Lead, South Dakota, USA, for an above ground run while final excavation and construction of the underground laboratory concluded. The surface laboratory was meant to mimic the underground environment to achieve as similar installation and operating conditions as possible. The same detector stand, the same transportation structures, and the same style of cleanroom served as excellent tools to perfect underground deployment. The only significant difference between the surface laboratory and the underground facility was a diminished surface water tank (3~m diameter) versus the subterranean tank (8~m diameter). Without the rock overburden of the underground lab, the surface run had significantly higher background rates than are allowable for any rare event dark matter search. The surface water tank provided observable particle flux moderation but did not provide anything similar to operating deep underground.

The surface run began on September 1, 2011 and ended February 14, 2012 after over 100~days of cryogenic detector operation. The LUX surface run accomplished all set goals and demonstrated all subsystems are capable of performing at or beyond operational goals~\cite{ak_run2}, with the exception of a single lose fitting that partially compromised our circulation path.

\subsection{Surface Results}

The xenon purification system achieved an electron lifetime of 204~$\pm$~6~$\muup$s while purifying at 35~slpm, as shown in Figure~\ref{fig:results} (left). This rate is equal to 300~kg/day (a majority of the LUX mass). We achieved $>$98\% heat exchanger efficiency ($<$5 W heat load). The light collection measurements indicated 8~phe/keVee ($>$4~phe/keVee) at the center of the detector with zero field (field-adjusted, scaled to 122 keV) with lower bounds on PTFE reflectivity of $>$95\%  and a photoabsorption length of $>$5~m.

LUX implemented the Mercury position reconstruction algorithm~\cite{so} which uses a maximum likelihood analysis of measured phototube light response functions and the measured intensity of the S2 signal in each PMT to estimate the ($x,y$) location of an event. Figure~\ref{fig:results} (right) shows the dodecagonal silhouette of the PTFE walls of the interior of LUX as well as the clear orientation of the thin gate grid wires that create the electron extraction field just below the liquid surface. The 5~mm spacing of the gate grid wires provide a relative indication of the performance of the reconstruction.

In addition, using the coincident decay of the 3.3~MeV $\beta^-$ and then the 7.7~MeV $\alpha$ radiation from decay chain  
${}^{214}$Bi 
$\rightarrow$
${}^{214}$Po 
$\rightarrow$
${}^{210}$Pb,
our reconstruction technique has shown to reconstruct with a resolution of $\sim$7~mm in each of the $x$ and $y$ directions. Although these are significantly higher energies than the dark matter energy window, a lower photomultiplier gain was required for surface running to prevent possible photocathode degradation due to large and continuous S2 signals from cosmic ray backgrounds, thus allowing the comparison of high energy, low gain S2 reconstruction to projected low energy, high gain S2 reconstruction.

\begin{figure}
\begin{center}
\begin{minipage}{0.45\linewidth}
\centerline{\includegraphics[width=1\linewidth,draft=false]{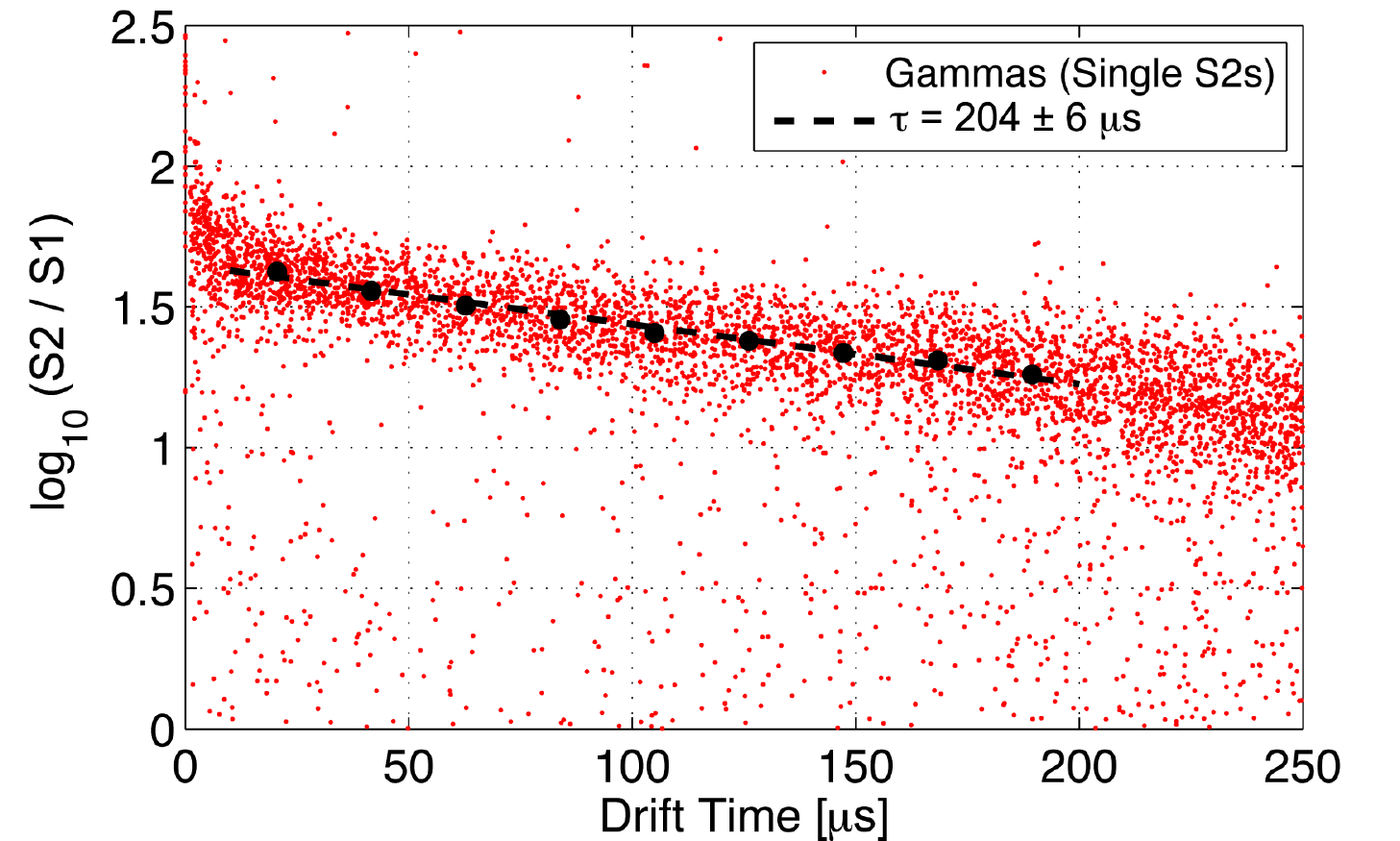}}
\end{minipage}
\begin{minipage}{0.45\linewidth}
\centerline{\includegraphics[width=1\linewidth,draft=false]{gate_position_reco.pdf}}
\end{minipage}
\caption{Measurement of the 204~$\pm$~6~$\muup$s drift time (left) and reconstruction of $x,y$ event locations using the Mercury reconstruction algorithm (right).}
\label{fig:results}
\end{center}
\end{figure}

\section{Commissioning}

Following the successful operation on the surface, we emptied the detector of xenon and transported the LUX detector 1480~m (4300~m.w.e) underground. The process of transporting a fully instrumented detector and attached breakout systems began on July 11, 2012. Two days of meticulous adjustments and translations culminated with LUX in the new Davis Laboratory ready for deployment. Unpacking and installation proceeded until September 2012. In October, the water tank was filled (Figure~\ref{fig:submerged}) and the detector began collecting vacuum  \v{C}erenkov data. With the introduction of gaseous xenon in December 2012, LUX began a stabilization and purification period to better understand run conditions and parameterize operating behaviors underground.

\begin{figure}
\begin{center}
\begin{minipage}{0.33\linewidth}
\centerline{\includegraphics[width=1.5\linewidth,draft=false]{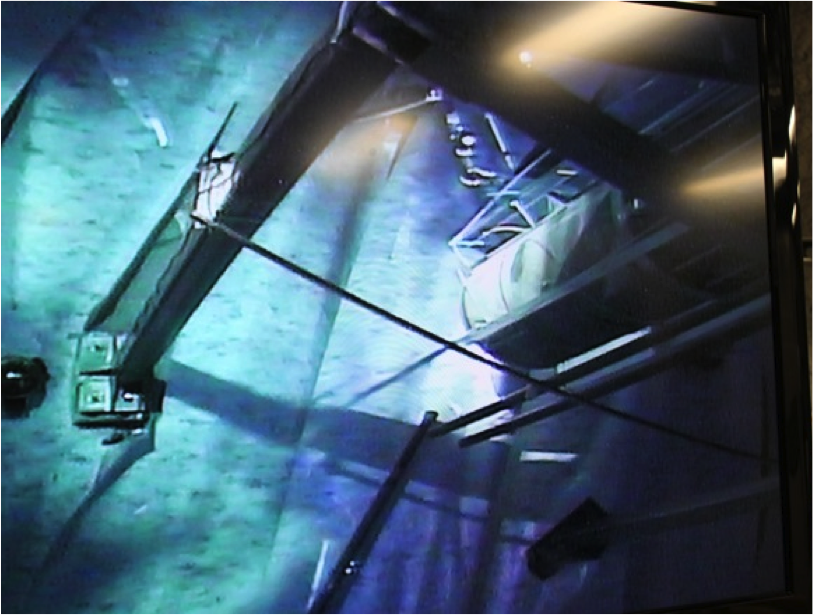}}
\end{minipage}
\caption{The LUX detector situated in the filled 8~m by 6~m  water tank. Installation occurred while the tank was empty of water and personnel. A portion of the detector stand is visible on the left, with the outer cryostat of the detector on the right.}
\label{fig:submerged}
\end{center}
\end{figure}

\section{Outlook}

The LUX detector was successfully deployed and is ramping up for a preliminary dark matter run in 2013. The detector is currently operating underground, and is circulating and purifying liquid xenon at SURF. Calibrations are underway to improve our understanding of the detector response in a quiet, underground environment. After full detector commission is finished and all subsystems have been operationally verified, we will begin a 60~day WIMP dark matter run with results expected by the end of the year.

\section*{Acknowledgements}
This work was partially supported by the U.S. Department of Energy
(DOE) under award numbers DE-FG02-08ER41549, DE-FG02-91ER40688, DOE,
DE-FG02-95ER40917, DE-FG02-91ER40674, DE-FG02-11ER41738,
DE-FG02-11ER41751, DE-AC52-07NA27344, the U.S. National Science
Foundation under award numbers PHY-0750671, PHY-0801536, PHY-1004661,
PHY-1102470, PHY-1003660, the Research Corporation grant RA0350, the
Center for Ultra-low Background Experiments in the Dakotas (CUBED),
and the South Dakota School of Mines and Technology
(SDSMT). LIP-Coimbra acknowledges funding from Funda\c{c}\~{a}o para a
Ci\^{e}ncia e Tecnologia (FCT) through the project-grant
CERN/FP/123610/2011.  We gratefully acknowledge the logistical and
technical support and the access to laboratory infrastructure provided
to us by the Sanford Underground Research Facility (SURF) and its
personnel at Lead, South Dakota.

\end{document}